\documentclass[pdflatex]{sn-jnl}

\usepackage{graphicx}%
\usepackage{multirow}%
\usepackage{amsmath,amssymb,amsfonts}%
\usepackage{amsthm}%
\usepackage{mathrsfs}%
\usepackage[title]{appendix}%
\usepackage{xcolor}%
\usepackage{textcomp}%
\usepackage{manyfoot}%
\usepackage{booktabs}%
\usepackage{algorithm}%
\usepackage{algorithmicx}%
\usepackage{algpseudocode}%
\usepackage{listings}%

\usepackage{tcolorbox}
\usepackage{natbib}
\newtcolorbox{promptbox}[2][]{
colback     = #2!5,
colframe    = #2!60!black,
fonttitle   = \bfseries,
fontupper   = \small,
title       = #1,
boxrule     = 0.4pt,
arc         = 3pt,
left        = 6pt,
right       = 6pt,
top         = 4pt,
bottom      = 4pt
}
\newcommand{\T}[1]{\texttt{#1}}
\newcommand{\B}[1]{\T{\textcolor{violet}{#1}}}
\newcommand{\V}[1]{\B{\{#1\}}}

\theoremstyle{thmstyleone}%
\theoremstyle{thmstyletwo}%

\theoremstyle{thmstylethree}%

\begin{document}

\title[Article Title]{LLM-based uncertainty assessment of social media situational signals for crisis reporting}

\author[1,3]{\fnm{Timothy} \sur{Douglas}}\email{timothy.douglas.21@ucl.ac.uk}

\author*[2,3]{\fnm{Roben} \sur{Delos Reyes}}

\author[3]{\fnm{Asanobu} \sur{Kitamoto}}

\affil*[1]{\orgdiv{Department of Computer Science}, \orgname{University College London}, \orgaddress{\city{London}, \country{United Kingdom}}}

\affil[2]{\orgdiv{School of Computing and Information Systems}, \orgname{The University of Melbourne}, \orgaddress{\street{Parkville}, \state{Victoria}, \country{Australia}}}

\affil[3]{\orgname{National Institute of Informatics}, \orgaddress{\city{Tokyo}, \country{Japan}}}


\abstract{Social media has become a critical source of situational awareness during disasters, providing real-time insights into evolving impacts and emerging needs. To support crisis response at scale, recent work has increasingly leveraged large language models (LLMs) to automatically classify and summarize situational information from social media streams. However, existing approaches implicitly assume that extracted situational claims are equally plausible, despite information quality varying substantially as a crisis unfolds. In this work, we propose an uncertainty-aware framework for automated situational awareness reporting that explicitly accounts for the plausibility of social media claims. First, we classify social media posts according to an established situational awareness schema. Second, we introduce an uncertainty assessment layer that evaluates whether individual situational claims plausibly reflect real-world conditions when conditioned on external proxy data, while explicitly eliciting the model's confidence in this judgment. Third, we use these uncertainty assessments to generate crisis reports that communicate not only what is being reported, but how certain those reports are. We apply this framework to over 200,000 earthquake-related Twitter/X posts, using impact summaries from the USGS PAGER as a representative external proxy. We argue that explicitly representing uncertainty supports human crisis communicators in prioritizing information under time pressure, and provides a framework for integrating external proxy data into LLM-based situational awareness pipelines.}


\maketitle
\section{Introduction}\label{intro}
Social media has become a critical source of situational awareness during disasters, offering rapid, fine-grained insights into evolving impacts, emerging needs, and human behavioral responses. Crisis informatics research has repeatedly shown that platforms such as Twitter/X (hereafter, X) can support spatiotemporal awareness in terms of who is affected, where impacts are occurring, and how conditions change over time. As a result, there has been a strong push to automate the extraction and aggregation of situational signals from social media at scale, in order to support timely decision-making during emergencies.

Recent advances in large language models (LLMs) have further accelerated this trend. Studies have found that LLM-based systems can classify crisis-related social media content and generate situational summaries and reports quickly and at low cost, making them attractive for real-time deployment. However, most existing approaches implicitly assume that situational signals extracted from social media can be treated as plausible inputs to downstream reports once they have been classified \cite{cantini2025harnessing, belcastro2025multistakeholder}. This assumption does not hold in practice. Prior work has consistently shown that social media reports during unfolding disasters can be incomplete, contradictory, and difficult to verify, with information quality varying across time, geography, and source \cite{shim2020quality, imran2015processing}.

This creates a critical gap for operational crisis intelligence. Decision-makers not only need to know what is being reported, but also how plausible those reports appear given available information. Treating all situational signals as equally plausible risks amplifying noise or discarding meaningful signals simply because they deviate from expected patterns. Existing automated pipelines rarely communicate this uncertainty downstream \cite{cantini2025harnessing, belcastro2025multistakeholder}, potentially limiting their usefulness in real-time settings.

In this work, we address this gap by introducing a framework for assessing the uncertainty of social media situational signals using external proxy data. The framework conditions an LLM on both an individual social media post and event-level proxy data (e.g., impact estimates) and asks whether the situational signal plausibly reflects those broader real-world conditions. Rather than serving as additional prompt context, the proxy provides independently collected, event-level impact data that grounds plausibility judgments in real-world conditions. The model is therefore prompted to reason over both social media metadata and structured external proxy data when forming its assessment.

A central contribution of our approach is that we elicit not only a plausibility judgment for each signal, but also the model's expressed confidence in that judgment. Recent studies suggest that LLMs can self-express their confidence when prompted appropriately \cite{naderi2025evaluating, tian2023just, xiong2023can}. Building on this, we elicit both a graded plausibility score and a confidence estimate for each social media signal. This enables downstream systems to distinguish between expected, high-confidence signals; implausible or noisy content; and uncertain but potentially important anomalies that might require further verification.

We empirically evaluate this framework through a case study of six earthquake events, using over 200,000 earthquake-related X posts (hereafter, tweets) collected in prior work \cite{li2023exploring, li2021social}. We use impact summaries from the USGS PAGER system as a representative real-world proxy of earthquake impact. This allows us to examine how plausibility and model confidence vary across events.

Finally, we demonstrate how proxy-informed plausibility and model confidence estimates can be operationalized for crisis reporting. By structuring reports according to different levels of plausibility and confidence, the framework enables triage under time pressure. Instead of producing a single undifferentiated report, the system generates multiple reports corresponding to different uncertainty profiles. This paper addresses the following research questions:

\begin{itemize}
    \item \textbf{RQ1.} Does the uncertainty-aware framework produce variation in plausibility and confidence estimates over situational signals when conditioned on external proxy information?
    \item \textbf{RQ2.} Does conditioning report generation on these uncertainty estimates measurably alter report structure compared to a baseline that treats all signals as equally plausible?
\end{itemize}

To address these questions, this paper makes three main contributions:

\begin{enumerate}
    \item We introduce a novel framework for proxy-informed uncertainty assessment of social media situational signals using LLMs.
    \item We provide an empirical evaluation of this framework using over 200,000 tweets related to six earthquake events varying in severity and scale.
    \item We demonstrate how plausibility and confidence estimates can be operationalized to generate differentiated crisis reports that help analysts prioritize information under time pressure.
\end{enumerate}

The remainder of this paper is organized as follows. We begin with a review of relevant literature (Section \ref{sec:related_work}). We then address our first contribution by describing the proposed uncertainty-aware reporting framework, including situational awareness classification, proxy-informed uncertainty assessment, and crisis reporting (Section \ref{sec:methodology}). Next, we address our second contribution by performing an empirical analysis of our methodology on 200,000 earthquake-related tweets (Section \ref{sec:case_study}). We subsequently address our third contribution by demonstrating how uncertainty-assigned tweets can be operationalized for crisis reporting (Section \ref{sec:results}). Finally, we discuss our findings (Section \ref{sec:discussion}), outline the limitations of the proposed approach (Section \ref{sec:limitations}), and conclude (Section \ref{sec:conclusion}).
\section{Related work}\label{sec:related_work}

\subsection{Situational awareness during crisis}

Prior work has extensively highlighted the importance of situational awareness (SA) for disaster management. SA has been defined as an ``idealized state of understanding what is happening in an event with many actors and other moving parts, especially with respect to the needs of command and control operations'' \cite{vieweg2010microblogging}. As such, SA enables responders to move beyond simply knowing \emph{that} an event is occurring toward understanding \emph{what} is happening, \emph{where} it is happening, and \emph{how} conditions are evolving \cite{wang2019space}. This geographically-grounded understanding of real-time developments during disasters is essential for informed decision-making, resource allocation, and response coordination \cite{maceachren2011senseplace2, huang2015geographic}.

Given the increasing role of digital communication during disasters, social media platforms have emerged as a valuable source of situational information \cite{wang2019space, imran2015processing}. Platforms such as X enable affected populations to share real-time observations about their immediate environment, including reports of damage, requests for assistance, and descriptions of ongoing response activities \cite{imran2013extracting, vieweg2010microblogging}. Recent work has demonstrated that social media can support SA across all phases of disaster management, from preparedness and early warning through emergency response and long-term recovery \cite{zade2018situational}.

However, extracting actionable situational information from social media at scale remains challenging. The volume of content generated during disasters far exceeds what human analysts can manually process. This has motivated growing interest in computational methods that summarize raw social media streams into interpretable reports.

\subsection{Automated crisis reporting}\label{sec:related_work_auto}

Building on this motivation, a growing body of research focuses on automated crisis reporting, which aims to transform classified social media posts into coherent situation reports that can support decision-making under time pressure. Recent advances in large language models (LLMs) have accelerated progress in this area, given their capabilities in generating coherent text summaries. Previous studies have illustrated this trend. Seeberger and Riedhammer \cite{seeberger2024crisis2sum}, for example, apply instruction-based prompting with LLMs to generate multi-query focused disaster summaries. Belcastro \textit{et al.} \cite{belcastro2025multistakeholder} combine BERT-based classification with generative models such as ChatGPT to produce human-readable reports tailored to distinct stakeholder groups. Gonella \textit{et al.} \cite{gonella2025crisitext} adopt a similar generative approach, using GPT-4o-mini to produce warning messages designed to guide civilians during and after crisis events. Most closely related to the present work, Cantini \textit{et al.} \cite{cantini2025harnessing} propose a framework that leverages prompt-based LLMs to monitor social media and generate automated disaster reports by aggregating citizen-reported issues. Collectively, these studies demonstrate the potential of LLMs to deliver timely, cost-effective, and increasingly sophisticated situational summaries.

Despite these advances, many automated reporting frameworks implicitly treat classified social media posts as equally plausible inputs once they have been assigned to a situational awareness category (e.g., Cantini \textit{et al.} \cite{cantini2025harnessing} and Belcastro \textit{et al.} \cite{belcastro2025multistakeholder}). Even where filtering mechanisms are used to select relevant content, the underlying claims themselves are not assessed for plausibility relative to real-world conditions. The implicit assumption is that classifying tweets and filtering them accordingly is sufficient to identify plausible situational claims.

However, this assumption presents a challenge in operational settings. For example, during the acute phases of a disaster, information can be highly varied. Situational claims made in early reports are often fragmented, exaggerated, or contradictory. Treating a verified observation from an official source with the same level of certainty as an unverifiable or exaggerated claim from an anonymous account risks producing misleading summaries. Moreover, because the plausibility of social media content can vary substantially across time, location, and source, pipelines that do not differentiate among claims may inadvertently amplify noise or obscure weak but genuine signals that diverge from expected patterns. Addressing this limitation requires mechanisms for generating reports that communicate not only what is being reported, but also the level of confidence associated with those reports.

\subsection{Uncertainty assessment}

While automated crisis reporting systems demonstrate the generative capabilities of LLMs, they typically do not model uncertainty inherent to claims made on social media. Addressing this limitation requires mechanisms for plausibility assessment and explicit uncertainty expression. A complementary body of literature on confidence elicitation and graded judgment in LLMs provides methodological foundations for this task. Here, we use the term \textit{epistemic} uncertainty to refer to uncertainty arising from incomplete knowledge about the plausibility of a situational claim, given available social media content and external proxy data.

As LLMs are increasingly deployed in high-stakes settings, the ability to accurately express confidence in their own outputs has become critical for trustworthy decision-making \cite{xiong2023can, tian2023just}. Xiong \textit{et al.} \cite{xiong2023can} provide a systematic framework for confidence elicitation in LLMs, identifying three core components: prompting strategies that instruct models to verbalize their confidence; sampling strategies that generate multiple responses to capture consistency; and aggregation strategies that combine these signals into uncertainty estimates. Their empirical evaluation demonstrates that approaches combining verbalized confidence with consistency-based methods yield the strongest results. This framework has since been extended across domains including medical question answering \cite{naderi2025evaluating} and long-form generation \cite{huang2024calibrating}, where confidence must be assessed for partially correct responses rather than binary outcomes.

A related strand of work has explored Likert-scale prompting as a mechanism for eliciting graded judgments from LLMs. Adapted from survey methodology, Likert scales enable models to express assessments along a continuum rather than forcing binary decisions \cite{plebe2025ll, lee2024chatfive}. This approach has been used to evaluate answer correctness, helpfulness, bias, harmfulness, and other aspects of model behavior or personality. For plausibility assessment, where the relationship between a social media claim and available contextual evidence is rarely concrete, this graded formulation offers a more expressive alternative. Framing plausibility as a construct ranging from highly plausible to implausible allows for granular reasoning about the degree to which a claim could plausibly reflect real-world conditions.

Our framework operationalizes plausibility assessment as a structured reasoning task that combines graded judgment with confidence elicitation. For each classified tweet, the model produces a plausibility score on a five-point Likert scale alongside a separate confidence estimate reflecting its certainty in that judgment. This framework adapts the methodology developed in Xiong \textit{et al.}, \cite{xiong2023can} while drawing on other established techniques for eliciting human-like judgments from LLMs \cite{plebe2025ll}. The uncertainty assessment step transforms raw situational signals into tweets with explicit plausibility and confidence estimates, producing reports that help crisis responders prioritize actionable information under time pressure.
\section{Methodology}\label{sec:methodology}

\begin{figure}[t!]
    \centering
    \includegraphics[width=11cm]{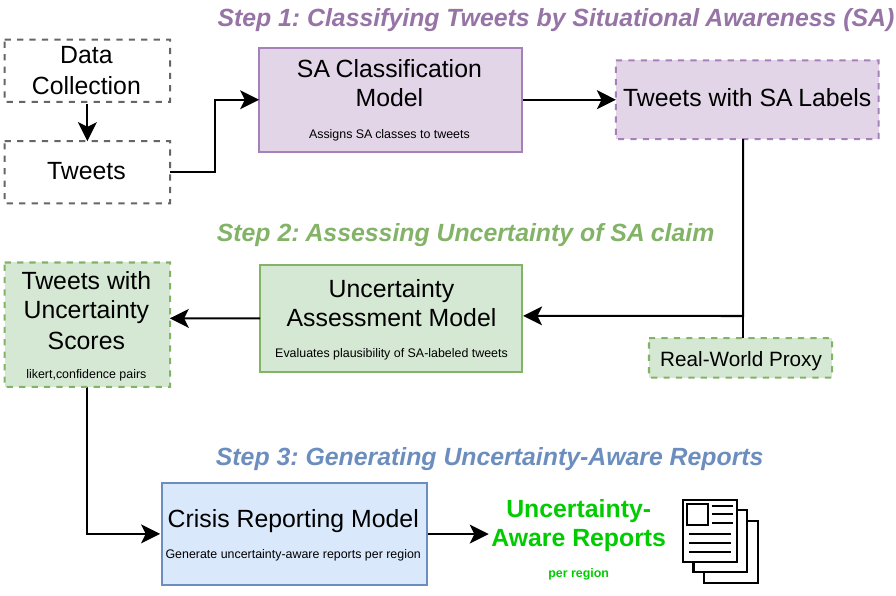}
    \caption{Overview of the proposed uncertainty-aware report generation framework.}
    \label{fig:method}
\end{figure}

This section outlines our framework for assessing the uncertainty of social media situational signals using external proxy data. The pipeline comprises a data collection and pre-processing module followed by three key steps: (i) classification, (ii) uncertainty assessment, and (iii) report generation. While classification and report generation largely build on existing approaches, the uncertainty assessment step constitutes the core methodological contribution of this work. The framework is designed to be modular and applicable to different types of crisis case studies. A summary of our methodology is illustrated in Figure \ref{fig:method}.

\subsection{Data collection}

We use X (previously Twitter) as the source of social media data in this study. X is among the most popular micro-blogging platforms, with 251 million daily active users as of the second quarter of 2024 \citep{twitter2024}. X has been widely adopted in disaster research due to its global reach and near-real-time nature, making it particularly suitable for situational awareness analysis during rapidly unfolding events.

We acknowledge that access to X data has become increasingly restricted in recent years, raising concerns about long-term reproducibility and scalability \cite{reyes2026}. However, the proposed framework is platform-agnostic. It requires only two inputs: (i) short-form, user-generated textual reports (e.g., social media posts or crowdsourced observations), and (ii) external real-world proxy data summarized into a form suitable for zero-shot LLM prompting. As such, the approach can potentially be applied to alternative social media platforms or other crowdsourced data sources where similar uncertainty challenges arise. This abstraction further enables the use of synthetic or simulated social media streams as inputs (for example, see \cite{reyes2026}), supporting evaluation under constrained or unavailable real-world data conditions.

\subsection{Step 1: Classification of situational awareness}

In the first step of our pipeline, crisis-related tweets are assigned one or more situational awareness (SA) categories using an existing SA schema. This step is necessary to ensure we assess uncertainty based on distinct situational claims extracted from tweets (e.g., infrastructure damage or impacts on people), as opposed to generic claims. Organizing tweets according to an established SA schema enables downstream aggregation and comparison across claim types, while preserving interpretability for operational users. The framework does not prescribe a specific taxonomy; instead, classification is treated as a modular component that can accommodate different schemas depending on the crisis context and reporting needs.

We implement this classification step using zero-shot LLM prompting (see Box 1). We design this as a multi-label classification task, allowing the model to assign multiple categories to a single post. This decision was motivated by the belief that individual tweets can convey several situational cues (for example, both affected people and infrastructure damage), in line with prior work \cite{douglas2026tracking, choi2020ten}. The output of this step is the original tweet dataset, now enriched with one or more SA category labels assigned to each tweet. The purple boxes in Figure \ref{fig:method} summarize this step.

\begin{promptbox}[Box 1: Classification prompt]{gray}
\textbf{System role.}  
You are an information extraction system assisting crisis impact researchers.
\\
\textbf{Task.}  
Classify each tweet into one or more situational awareness categories and extract the key text spans (``rationales'') that justify each classification.
\\
\textbf{Situational awareness schema.}
\\
\V{SITUATIONAL AWARENESS SCHEMA}
\\
\textbf{Input.}  
Now read the following social media posts:
\\
\V{SOCIAL MEDIA POSTS}
\end{promptbox}

\subsection{Step 2: Uncertainty assessment}

While SA classification identifies the \textit{type} of situational signal a tweet contains, it offers limited insight into whether that signal is plausible given real-world conditions. We define \textit{plausibility} as the degree to which a situational signal is consistent with independently available event-level evidence. We define \textit{confidence} as the model's self-expressed certainty in its own plausibility judgment. These two concepts address distinct sources of uncertainty. The first reflects epistemic uncertainty in the tweet claim itself (arising from incomplete, delayed, or conflicting information about evolving conditions), while the second captures the model's uncertainty in its assessment of plausibility. Explicitly modeling both enables downstream reports to distinguish between strongly supported signals (e.g., high plausibility and confidence) and more tentative observations (e.g., low plausibility and confidence), supporting risk-aware decision-making under time pressure.

To address this, we design a proxy-informed uncertainty assessment module that evaluates the plausibility of situational claims made in tweets. Let $\mathcal{T} = \{t_1, t_2, \dots, t_N\}$ denote a set of crisis-related tweets. Each tweet is represented as follows:
\begin{equation}
    t_i = (x_i, s_i, \tau_i, y_i),
    \label{eq:tweets}
\end{equation}
where $x_i$ is the textual component, $s_i$ is spatial metadata, $\tau_i$ is the timestamp, and $y_i \subseteq \mathcal{Y}$ denotes the assigned situational awareness (SA) label(s) drawn from the SA schema $\mathcal{Y}$.

In parallel, we incorporate structured external proxy data $\mathcal{P}$ describing independently collected impact estimates. The proxy layer provides contextual evidence about expected conditions across affected locations. In our formulation, the proxy is spatially indexed and static for a given event. Plausibility assessment therefore conditions each tweet $t_i$ on a location-specific subset of the proxy, denoted:
\begin{equation}
    P_i = \mathcal{P}(s_i),
    \label{eq:proxy}
\end{equation}
rather than on a time-conditioned proxy representation. The model is prompted to assess the plausibility of each situational claim by reasoning jointly over (i) the tweet representation $t_i$ and (ii) proxy-derived impact summary $P_i$. 

For each tweet $t_i \in \mathcal{T}$ and its corresponding location-conditioned proxy summary $P_i$, the uncertainty assessment module produces a plausibility score $l_i$ and confidence score $c_i$:
\begin{equation}
    (l_i, c_i) = f(t_i, P_i),
    \label{eq:uncertainty}
\end{equation}
where $l_i \in \{1,2,3,4,5\}$ is a graded plausibility score, and $c_i \in [0,100]$ is a confidence score expressing certainty in that plausibility judgment. 

\subsubsection{Real-world proxy data}\label{sec:methodology_proxy}

This subsection describes the proxy representation $P_i$ defined in Equation \ref{eq:proxy}. Assessing the plausibility of a situational claim requires reference to information beyond the claim itself. Social media posts can vary in their plausibility, particularly during the early stages of a crisis. Without external context, plausibility judgments risk collapsing into purely linguistic assessments separated from conditions specific to an event. To mitigate this, we condition plausibility reasoning on structured real-world proxy data describing independently collected impact estimates for the relevant crisis.

The role of the proxy is to provide event-level contextual anchors that enable structured reasoning about whether a claim could reasonably reflect observed conditions. It is not treated as ground truth, nor is the objective to enforce agreement between tweet content and proxy impact estimates. A claim may legitimately diverge from the proxy due to localized effects, delayed reporting, or other credible contextual cues. Conversely, superficial alignment with proxy summaries does not eliminate uncertainty.

To support this uncertainty reasoning process, proxy data must satisfy three criteria. First, it must be spatially aligned with the social media signals under analysis, ensuring that plausibility judgments are conditioned on comparable contexts. Second, it must contain interpretable and quantifiable indicators of event impact (e.g., severity estimates or exposure summaries) that can inform reasoning about expected conditions. Third, it must be available in a structured format so that summaries can be programmatically incorporated into LLM prompts. These requirements ensure that proxy data can be integrated to strengthen contextual grounding in the uncertainty assessment task.

\subsubsection{Uncertainty assessment by plausibility and confidence}\label{sec:methodology_uncertainty}

Given tweet $t_i$ and proxy information $P_i$ (described in Section \ref{sec:methodology_proxy}), the uncertainty assessment stage estimates two quantities defined in Equation \ref{eq:uncertainty}: the plausibility score $l_i$ and the confidence score $c_i$.

\paragraph{Definition of plausibility and confidence}

The core idea of our uncertainty assessment stage is to represent each situational claim using two complementary dimensions --- plausibility and confidence. Together, these quantities define a two-dimensional epistemic profile for each tweet and operationalize uncertainty within the proposed framework. 

\begin{itemize}
    \item Plausibility $(l_i)$ refers to the degree to which a tweet's situational claim is reasonable given the available contextual evidence, including event metadata and structured proxy-derived impact summaries. It does not assess factual correctness; rather, it captures contextual alignment with expected real-world conditions.
    \item Confidence $(c_i)$ refers to the model's self-expressed certainty in its plausibility judgment. Low confidence indicates that the model perceives the claim as ambiguous or under-supported by contextual evidence.
\end{itemize}

We adopt a graded, rather than binary, representation of plausibility for two reasons. First, crisis-related claims may not be verifiably true or false at the time they are posted. A claim may be broadly consistent with event severity yet lack key details, or may diverge from aggregate patterns while remaining locally credible. A five-point Likert scale ($1=\text{implausible}$, $5=\text{highly plausible}$) enables the model to express intermediate states such as ambiguity or partial alignment. Prior work suggests that Likert-style prompting supports more expressive reasoning in LLMs \cite{plebe2025ll}.

\begin{promptbox}[Box 2: Uncertainty assessment prompt]{gray}
\textbf{System role.}  
You are an expert crisis analysis assistant.
\\
\textbf{Task.}  
Evaluate whether social media situational awareness (SA) signals
are plausibly representative of real-world conditions during a crisis event, given automatically generated impact estimates and contextual uncertainty.
\\

\textbf{Context.}
\\
Region: \V{MMI}
\\
...
\\
\textbf{External Proxy Data.}
\\
Proxy Summary: \V{SUMMARY}
\\
\textbf{Input.}  
Now read the following social media posts:
\\
\V{SOCIAL MEDIA POSTS}
\\
\textbf{Uncertainty assessment task.}  
For \emph{each tweet}, determine the following:

\begin{enumerate}
    \item \textbf{Plausibility (Likert 1–5).}  
    \begin{itemize}
        \item 5 = highly plausible
        \item ...
    \end{itemize}
    \item \textbf{Confidence (0–100).}  
    How confident are you in \emph{your plausibility judgment}?  
\end{enumerate}
\end{promptbox}

\paragraph{Uncertainty estimation strategies}

\begin{figure}[t!]
    \centering
    \includegraphics[width=13cm]{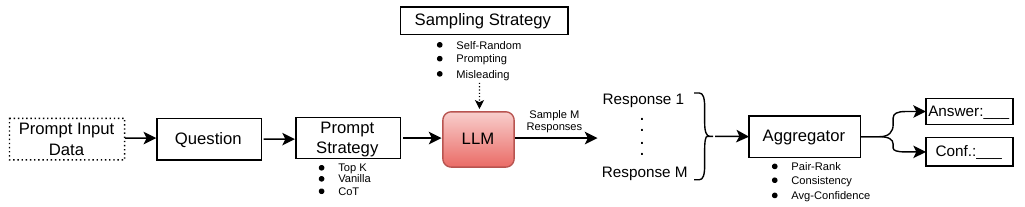}
    \caption{Overview of the confidence elicitation framework (adapted from Xiong \textit{et al.} \cite{xiong2023can}).}
    \label{fig:conf_method}
\end{figure}

To estimate plausibility and confidence, we adapt the confidence elicitation framework proposed by Xiong \textit{et al. } \cite{xiong2023can}, which decomposes uncertainty estimation into three components: (i) a prompting strategy, (ii) a sampling strategy, and (iii) an aggregation strategy (see Figure \ref{fig:conf_method}). This framework is particularly suitable for crisis contexts, where situational claims may be ambiguous, evolving, and difficult to verify in real time. Deterministic single-pass prompting risks over-committing to unstable judgments, whereas sampling-based elicitation exposes variability in reasoning and enables explicit uncertainty expression.

\subparagraph{Prompting strategy}
For each tweet $t_i$, we construct a structured prompt that presents: (i) the tweet text and metadata, (ii) the assigned SA label(s), and (iii) summarized proxy-derived impact indicators $P_i$. The model is instructed to jointly evaluate this information and return both a graded plausibility score and an explicit confidence estimate (see Box 2). Joint elicitation mirrors expert reasoning, where assessments are accompanied by expressions of certainty.

\subparagraph{Sampling strategy}
Following Xiong \textit{et al.} \cite{xiong2023can}, we obtain multiple independent generations per tweet using stochastic decoding. Sampling serves two purposes: (i) reducing sensitivity to any single model output, and (ii) surfacing variability in reasoning trajectories. In rapidly evolving crises, disagreement across sampled generations may reflect genuine contextual ambiguity rather than model error. Sampling therefore provides a mechanism for exposing epistemic fragility in situational claims. Let $(l_i^{(k)}, c_i^{(k)})$ denote the plausibility and confidence outputs from sample $k$.

\subparagraph{Aggregation strategy}
Sample-level outputs are aggregated to produce stabilized estimates:
\begin{equation}
l_i = \text{Agg}_l(\{l_i^{(k)}\}_{k=1}^{K}), 
\quad 
c_i = \text{Agg}_c(\{c_i^{(k)}\}_{k=1}^{K}).
\label{eq:agg}
\end{equation}
The specific aggregation and sampling procedures used in our experiments are described in Section \ref{sec:case_study}. 

Each situational claim is therefore represented by a two-dimensional epistemic profile defined by the pair $(l_i, c_i)$, where $l_i \in \{1,2,3,4,5\}$ denotes the plausibility of the claim given proxy-derived evidence and $c_i \in [0,100]$ denotes the model's expressed confidence in that plausibility assessment. Plausibility and confidence capture complementary aspects of uncertainty: a claim may receive a moderate plausibility score with high confidence, indicating a stable assessment of ambiguity, or a high plausibility score with low confidence, indicating sensitivity to limited or conflicting contextual evidence. These estimates form the basis for downstream operationalization and uncertainty-aware report generation.

\subsection{Step 3: Report generation}

The final stage of the pipeline aggregates uncertainty-aware situational claims into structured crisis reports. The uncertainty assessment stage produces, for each tweet $t_i$, a plausibility score $l_i$ and a confidence score $c_i$. These estimates are retained as continuous signals and inform downstream aggregation and summarization. Box 3 describes the prompt we use.

Automated crisis reporting is a common operational task in situational awareness analysis, where classified social media posts are synthesized into structured summaries for decision-makers. However, prior approaches typically treat all classified posts as equally plausible and certain once assigned a situational label (e.g., \cite{cantini2025harnessing}). This implicit assumption collapses variability in information quality and does not distinguish between strongly supported signals and more uncertain or ambiguous claims. In contrast, our framework preserves uncertainty estimates throughout the aggregation process. Rather than assuming all situational claims contribute equally to a summary, report generation can condition on plausibility and confidence signals when selecting and organizing content. This enables differentiation between highly plausible and well-supported claims, partially aligned or ambiguous information, and lower-plausibility signals, without requiring binary judgments of correctness.

As a result, generated reports communicate not only what situational information is present, but also reflect structured variation in informational status. This design supports more transparent and risk-aware crisis communication, where uncertainty is treated as an explicit component of situational awareness rather than an implicit artifact of model behavior.

\begin{promptbox}[Box 3: Report generation prompt]{gray}
\textbf{System role.}  
You are an AI system generating situation reports for crisis responders.
\\
\textbf{Task.}  
Your task is to support responder decision-making by clearly reporting what is being claimed and where uncertainty exists.
\\
\textbf{Input.}
\\
Event: \V{EVENT}
\\
Location: \V{GRID CELL ID}
\\
Number of Tweets: \V{NUMBER OF TWEETS}
\end{promptbox}

\subsection{Evaluation}

Evaluating uncertainty-aware situational awareness systems presents unique challenges, as ground-truth labels for plausibility and confidence are typically unavailable in real-time crisis settings. Rather than assessing factual correctness, our evaluation examines whether the proposed framework (i) introduces structured heterogeneity in situational claims and (ii) measurably alters downstream report composition. We therefore evaluate the pipeline with respect to the two research questions introduced in Section \ref{intro}, which address different stages of the system:

\begin{itemize}
    \item \textbf{RQ1.} examines whether uncertainty assessment introduces measurable variation among situational claims that would otherwise be treated uniformly.
    \item \textbf{RQ2.} examines whether this structured variation propagates into generated crisis reports in a meaningful way.
\end{itemize}

\paragraph{RQ1: Uncertainty distribution over situational signals}

This question asks whether the uncertainty-aware framework produces distributions of plausibility and confidence across situational claims. Under a baseline assumption commonly adopted in prior report generation pipelines, all classified tweets are implicitly treated as equally plausible and equally certain. Formally, this corresponds to assigning every tweet the same point, yielding a delta distribution with zero dispersion.

In contrast, our framework assigns each tweet $t_i$ a pair $(l_i, c_i)$, where $l_i \in \{1,2,3,4,5\}$ and $c_i \in [0,100]$. Let
\[
L = \{l_1, \dots, l_N\}, 
\quad
C = \{c_1, \dots, c_N\}
\]
denote the empirical distributions of plausibility and confidence across $N$ tweets. We evaluate the induced uncertainty structure using two complementary analyses.

\emph{(1) Empirical uncertainty distributions.}
We visualize the empirical distributions $p(l = k)$ and $p(c \in b)$, where confidence values are discretized into bins $b$ for interpretability. These distributions reveal how uncertainty mass is distributed across tweets and whether the model differentiates between situational signals.

\emph{(2) Entropy of uncertainty distributions.}
To quantify heterogeneity, we compute the Shannon entropy of plausibility and confidence:
\begin{equation}
\begin{aligned}
H(L) &= - \sum_{k=1}^{5} p(l = k) \log_2 p(l = k), \\
H(C) &= - \sum_{b} p(c \in b) \log_2 p(c \in b).
\end{aligned}
\label{eq:entropy}
\end{equation}
Higher entropy indicates greater dispersion of uncertainty estimates, whereas zero entropy corresponds to a baseline in which all tweets are treated as equally plausible. Entropy therefore provides a scalar measure of whether uncertainty assessment introduces structured variation into the situational claim space.

\paragraph{RQ2: Impact of uncertainty on generated reports}

RQ2 examines whether structured variation in $(l_i, c_i)$ is reflected in the informational composition of generated crisis reports. Rather than evaluating factual correctness, this analysis investigates whether conditioning report generation on uncertainty estimates produces measurable differences in semantic structure.

To examine this, we construct reports from subsets of tweets selected according to their uncertainty profiles. These subsets represent distinct regions of the plausibility–confidence space. We compare these reports against baseline summaries generated from the full tweet set without uncertainty-aware conditioning. Our analysis focuses on informational structure and semantic organization rather than superficial linguistic properties such as readability or stylistic fluency. We assess report-level behavior using two complementary, descriptive analyses.

\emph{(1) Informational differentiation between reports.} We examine whether reports generated from differently conditioned tweet subsets exhibit measurable variation in semantic content. Let each report $R_i$ be represented by an embedding vector $\mathbf{e}_i$ obtained from a sentence-embedding model. The similarity between reports $R_i$ and $R_j$ is computed using cosine similarity:
\begin{equation}
S(R_i, R_j) = \frac{\mathbf{e}_i \cdot \mathbf{e}_j}{\|\mathbf{e}_i\| \, \|\mathbf{e}_j\|}.
\label{eq:report_similarity}
\end{equation}
This metric does not assume that reports conditioned on different uncertainty regimes must be highly dissimilar. In real crisis contexts, multiple subsets may still reference overlapping situational themes. Instead, cosine similarity provides a quantitative means to compare how semantic composition varies across conditioning strategies. Observed differences (or similarities) therefore characterize how uncertainty-aware structuring influences report content, without implying that one configuration is inherently superior.

\emph{(2) Internal semantic coherence.} We also examine the internal semantic coherence of each report. Let a report $R$ consist of $n$ sentences $\{s_1, \dots, s_n\}$ with corresponding embeddings $\mathbf{v}_1, \dots, \mathbf{v}_n$. Internal coherence is estimated as the average pairwise cosine similarity:
\begin{equation}
C(R) =
\frac{2}{n(n-1)}
\sum_{i<j}
\frac{\mathbf{v}_i \cdot \mathbf{v}_j}
{\|\mathbf{v}_i\| \, \|\mathbf{v}_j\|}.
\label{eq:report_internal_consistency}
\end{equation}
Higher values of $C(R)$ indicate greater semantic similarity among sentences within a report. However, coherence is not assumed to monotonically correspond to report quality. For example, reports derived from noisy or heterogeneous tweet subsets may exhibit lower internal coherence if they summarize diverse or conflicting signals. Conversely, reports generated from tightly clustered content may appear more semantically focused. This metric therefore serves as a structural diagnostic, allowing comparison of how uncertainty-conditioned subsets affect internal report organization.

Together, RQ1 and RQ2 evaluate whether the proposed framework (i) introduces structured variation in uncertainty estimates at the tweet level and (ii) yields observable differences in downstream report structure when these estimates are operationalized.
\section{Case Studies}\label{sec:case_study}

To demonstrate the framework's applicability to real-world disaster events, we apply the uncertainty-aware pipeline to six earthquake case studies. This section introduces the selected events, outlines the situational awareness (SA) schema used for classification, and describes the external proxy data used to contextualize plausibility assessments. Together, these components define the empirical setting in which the framework is evaluated.

\subsection{Selection of crisis events}

We focus on earthquake events as our representative crisis case study. Earthquakes are sudden-onset, potentially destructive disasters that disrupt infrastructure and daily life with little warning. In their immediate aftermath, affected populations frequently turn to social media to report damage, seek assistance, and share situational updates, generating large volumes of user content before authoritative information is fully available. This creates an environment where early reports can often appear fragmentary, speculative, or locally inconsistent \cite{zhou2025automated}. In such settings, decision-making that relies on social media must be able to distinguish between more and less plausible claims, as these assessments can influence how response efforts are prioritized and resources allocated.

Prior work in crisis informatics has frequently used earthquakes to study the relationship between social media signals and physical impact, in part due to the availability of external proxy data such as seismic intensity estimates and damage assessments \cite{li2023exploring, li2021social}. However, few studies explicitly compare such real-world indicators with individual social media claims to assess their plausibility. Signals observed on social media do not necessarily reflect underlying physical conditions, particularly in the early stages of a disaster. Earthquakes therefore provide a suitable setting for examining how external proxy data can be used to contextualize and evaluate the plausibility of situational signals online. We evaluate our framework across multiple earthquake events spanning different magnitudes, geographic contexts, and impact profiles (see Table \ref{tab:case_studies}).

Our experiments use a curated earthquake X corpus from the study by Li\textit{ et al.} \cite{li2023exploring, li2021social}, which has been employed in multiple peer-reviewed analyses of post-earthquake situational awareness. This allows us to situate our evaluation within an established empirical benchmark while extending prior work through explicit uncertainty modeling.

\begin{table}
\centering
\begin{tabular}{p{2.1cm}|p{1.5cm}|p{3.6cm}|p{2.6cm}|p{0.8cm}}
    \toprule
   Event & Start Date & Epicenter & Time & Mag. \\
    \midrule
   2014 Chile & 04/01/14 & Approximately 95 km northwest of Iquique, Chile & April 1, 11:46pm (8:46pm locally) & 8.2 \\
   2014 Napa & 08/24/14 & South of Napa and at the northwest of American Canyon, California, U.S.& August 24, 10:20am (3:20am locally) & 6.0 \\
   2015 Nepal & 04/25/15 & East of the Gorkha District at Barpak, Gorkha & April 25, 06:11am (11:56am locally) & 7.8 \\
   2019 Ridgecrest & 07/06/19 & Northeast of Ridgecrest, California, U.S. & July 6, 3:19am (July 5 at 8:19pm locally) & 7.1 \\
   2021 Fukushima & 02/13/21 & East coast of Hoshu Prefecture, Fukushima, Japan & February 13, 2:07pm (11:07pm, locally) & 7.1 \\
   2021 Haiti & 08/14/21 & West of the Haitian capital Port-au-Prince & August 14, 12:29pm (08:29am locally) & 7.2 \\
   \bottomrule
\end{tabular}
\caption{Summary of earthquake case studies considered in our analysis (adapted from Li \textit{et al.} \cite{li2023exploring}), including major characteristics and impacts of each event (times in UTC). \textit{Mag.} refers to Magnitude.}
\label{tab:case_studies}
\end{table}

\subsection{Selection of SA classification schema}

Having defined the crisis case study and data source, we next specify the situational awareness (SA) schema used to classify tweets. For the earthquake case study, we adopt an established classification schema based on the work of Wang and Ye \cite{wang2019space}, which itself builds on earlier crisis informatics frameworks proposed by Imran \textit{et al.} \cite{imran2013extracting}. The full set of categories and their descriptions are provided in Table \ref{tab:SA_schema}. This schema captures key dimensions of disaster-related situational information commonly expressed on social media, including impacts on people, damage to infrastructure, and weather conditions.

We make a minor adaptation to the original schema by explicitly including a \textit{Response/Recovery} category. This modification is motivated by the observation that, during earthquake events, a substantial proportion of social media content relates to emergency response actions, official interventions, and early recovery efforts, particularly beyond the immediate impact phase. While the original schema effectively represents impact-oriented and needs-based information, the addition of this category allows later-stage situational awareness signals to be represented more explicitly.

\begin{table}[t!]
\centering
\begin{tabular}{p{3.2cm}|p{9cm}}
    \toprule
   Category & Description \\
    \midrule
   Caution/Advice & Tweets referring to warnings, preparation, advice, and tips
   \\
   Affected People & Tweets referring to people trapped, injured, missing, and killed
   \\
   Infrastructure/Utilities & Tweets referring to infrastructure damage, services closure, built environment, and collapsed structure
   \\
   Needs \& Donations & Tweets referring to donations, volunteering, relief, and fundraising
   \\
    Weather \& Environment & Tweets referring to weather conditions and environment
   \\
    Response/Recovery & Mentions of emergency response, cleanup, repair, or official action
   \\
   \bottomrule
\end{tabular}
\caption{Summary of SA classification schema considered in our analysis (adapted from Wang and Ye \cite{wang2019space}).}
\label{tab:SA_schema}
\end{table}

The resulting set of categories aligns with widely adopted notions of situational awareness in the crisis informatics literature \cite{zahra2020automatic, rudra2018extracting, verma2011natural, vieweg2010microblogging}, encompassing perception of the environment (e.g., weather and environmental conditions), comprehension of impacts (e.g., affected people and infrastructure), and awareness of ongoing or planned actions (e.g., response and recovery activities). This particular schema is selected for three reasons. First, it is well grounded in related literature and has been applied successfully to real-world crisis events. Second, the categories correspond closely to information needs commonly identified in operational disaster response, making them suitable for report-oriented analysis. Third, the schema is sufficiently broad, capturing both immediate impact signals and later-stage response-related content observed in earthquake scenarios.

\begin{figure}
    \centering
    \includegraphics[width=11cm]{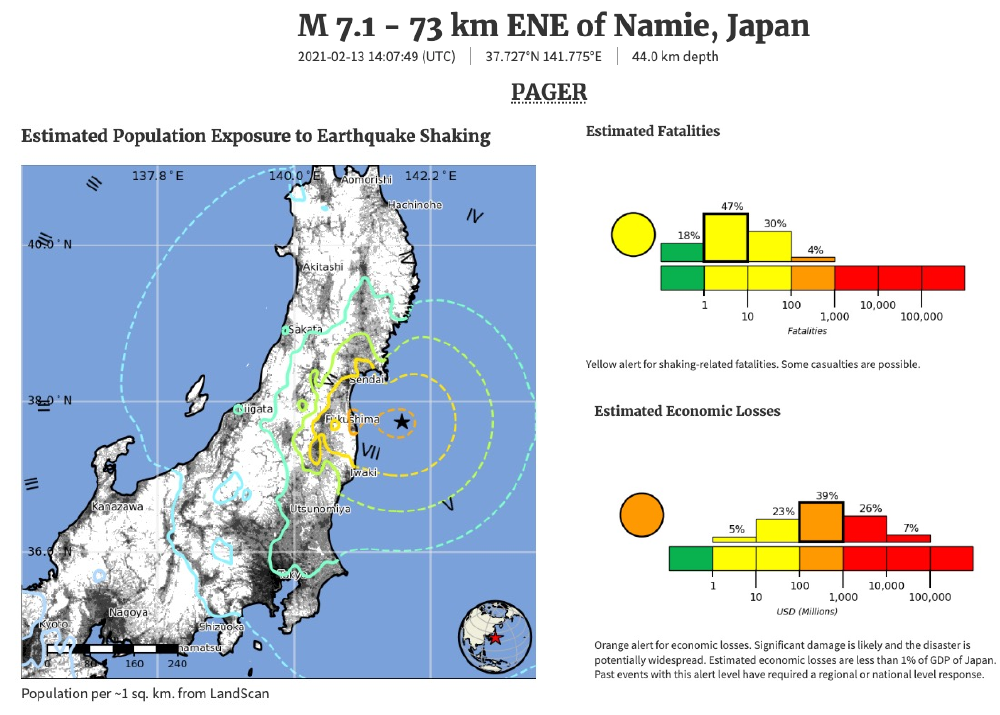}
    \caption{USGS PAGER summary for the 2021 Fukushima earthquake, including estimated fatalities and economic loss estimates.}
    \label{fig:pager}
\end{figure}

\subsection{Selection of real-world proxy data}

As outlined in Section \ref{sec:methodology_proxy}, our framework conditions plausibility reasoning on real-world proxy data describing independently collected impact estimates for a crisis. We now instantiate this proxy layer for the earthquake case study. We use earthquake impact estimates from the United States Geological Survey (USGS) Prompt Assessment of Global Earthquakes for Response (PAGER) system. PAGER provides near-real-time summaries of earthquake events, including Modified Mercalli Intensity (MMI) estimates, population exposure, and probabilistic estimates of fatalities and economic losses. A summary of this information is shown in Figure \ref{fig:pager}.

PAGER is well suited to our setting for two reasons. First, it provides structured, quantitative impact indicators that align spatially with the events under study. Second, our input datasets already include tweet-level mappings to Modified Mercalli Intensity (MMI) values derived from USGS ShakeMap data (following \cite{li2023exploring, li2021social}). This allows plausibility judgments to be conditioned directly on geographically grounded shaking intensity estimates. As a result, the proxy layer can provide event-level contextual evidence that is systematically linked to individual tweets.

\subsection{Quadrant-based operationalization of uncertainty}

Having instantiated the proxy layer and used this to obtain plausibility and confidence estimates for each tweet, we next consider how these outputs can be operationalized for downstream reporting. While the uncertainty assessment step (Section \ref{sec:methodology_uncertainty}) produces fine-grained plausibility and confidence scores $(l_i, c_i)$ for each tweet, evaluating their practical value requires structuring these outputs in a way that supports comparative report generation.

\begin{figure}[t!] 
    \centering 
    \includegraphics[width=8cm]{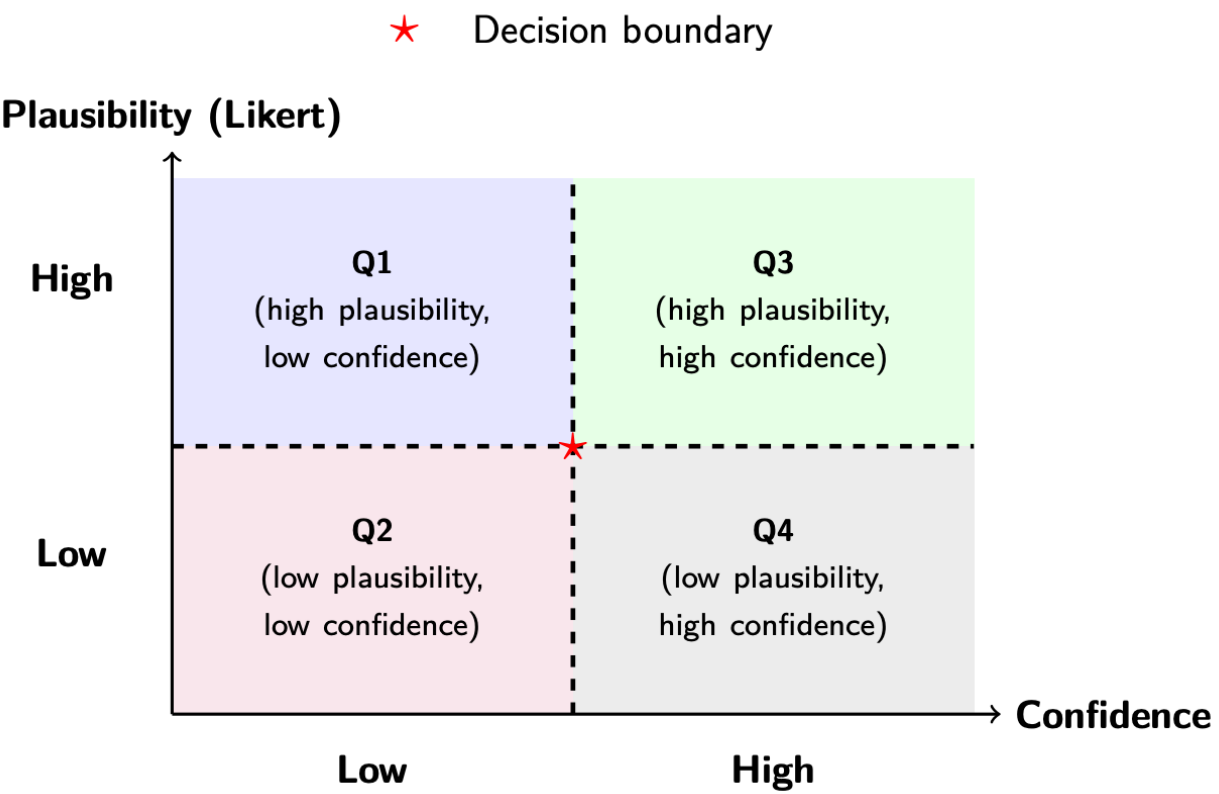} \caption{Operationalization of plausibility and confidence estimates into four regions used for report generation and evaluation.} \label{fig:quadrants} 
\end{figure}

As such, we partition the plausibility–confidence space into four regions defined by threshold values (Figure \ref{fig:quadrants}). This provides an intuitive way of grouping situational claims into distinct uncertainty regimes for report generation and comparative analysis. The purpose of this partitioning is not to insist that uncertainty is inherently categorical, nor to claim that these thresholds are optimal. Rather, the quadrant structure provides a transparent and interpretable instantiation that enables systematic experimentation.

This partitioning yields four regions corresponding to distinct uncertainty regimes: high plausibility, high confidence (Q3); high plausibility, low confidence (Q1); low plausibility, high confidence (Q4); low plausibility, low confidence (Q2). We emphasize that these regions do not imply normative judgments about correctness. Instead, they provide structured groupings that allow us to examine how reports differ when generated from tweets with distinct epistemic uncertainty profiles.

As shown in Figure \ref{fig:quadrants}, thresholds define boundaries between higher and lower plausibility and confidence values. In our experiments, we define $l=3$ (on a five-point Likert scale) as the plausibility boundary and $c=50$ (confidence midpoint) as the confidence boundary. These values represent neutral separation points rather than tuned decision thresholds. Alternative configurations could be adopted depending on risk tolerance or operational objectives.

\subsection{Experimental setup}

All experiments were conducted using the \texttt{google/gemma-2-9b-it} model deployed on an NVIDIA A100 GPU. The same model was used consistently across all components of the pipeline, including Step 1 (SA classification), Step 2 (uncertainty assessment), and Step 3 (report generation), ensuring consistency in the methodology and comparability of outputs. We selected \texttt{google/gemma-2-9b-it} instead of larger models because it is lightweight, enabling faster experimentation and iteration while remaining capable of the complex reasoning required for this task. Although larger or more recent models could be applied, evaluating model performance is not our primary contribution; our focus is on demonstrating the framework for uncertainty-aware situational reasoning and report generation.

During the uncertainty assessment stage, we instantiated the confidence elicitation framework using a vanilla prompting strategy (see \cite{xiong2023can} for more details). For each tweet, we generated $M=3$ independent samples using temperature values $\tau \in \{0.5, 1.0, 1.5\}$ to introduce controlled variability in model reasoning. These temperatures were selected to span low-, medium-, and high-variability regimes while remaining within reasonable ranges for stable text generation.

Furthermore, plausibility is elicited on a five-point Likert scale (1 = implausible, 5 = highly plausible). Although continuous scores could have been elicited, we deliberately chose a Likert scale to align model outputs with human-like evaluative reasoning. Likert-style responses have been recently used to elicit human judgments and personality traits in LLM research, enhancing the interpretability of model assessments (e.g., Plebe \textit{et al.} \cite{plebe2025ll}; Lee \textit{et al.} \cite{lee2024chatfive}). While our study does not aim to assess personality traits, Likert scales remain well suited for capturing graded, interpretable judgments that bridge machine reasoning and human decision-making. This approach might enable analysts to more intuitively interpret uncertainty judgments. Confidence scores, in contrast, were recorded as continuous values between 0 and 100, reflecting the model's self-assessed certainty.

Final plausibility and confidence values were computed by averaging across all samples. Specifically, plausibility scores were averaged and rounded to the nearest Likert level, while confidence scores were averaged directly to produce continuous estimates. All other generation parameters were held constant across experiments to ensure comparability.
\section{Results}\label{sec:results}

\begin{figure}[t!] 
    \centering 
    \includegraphics[width=13cm]{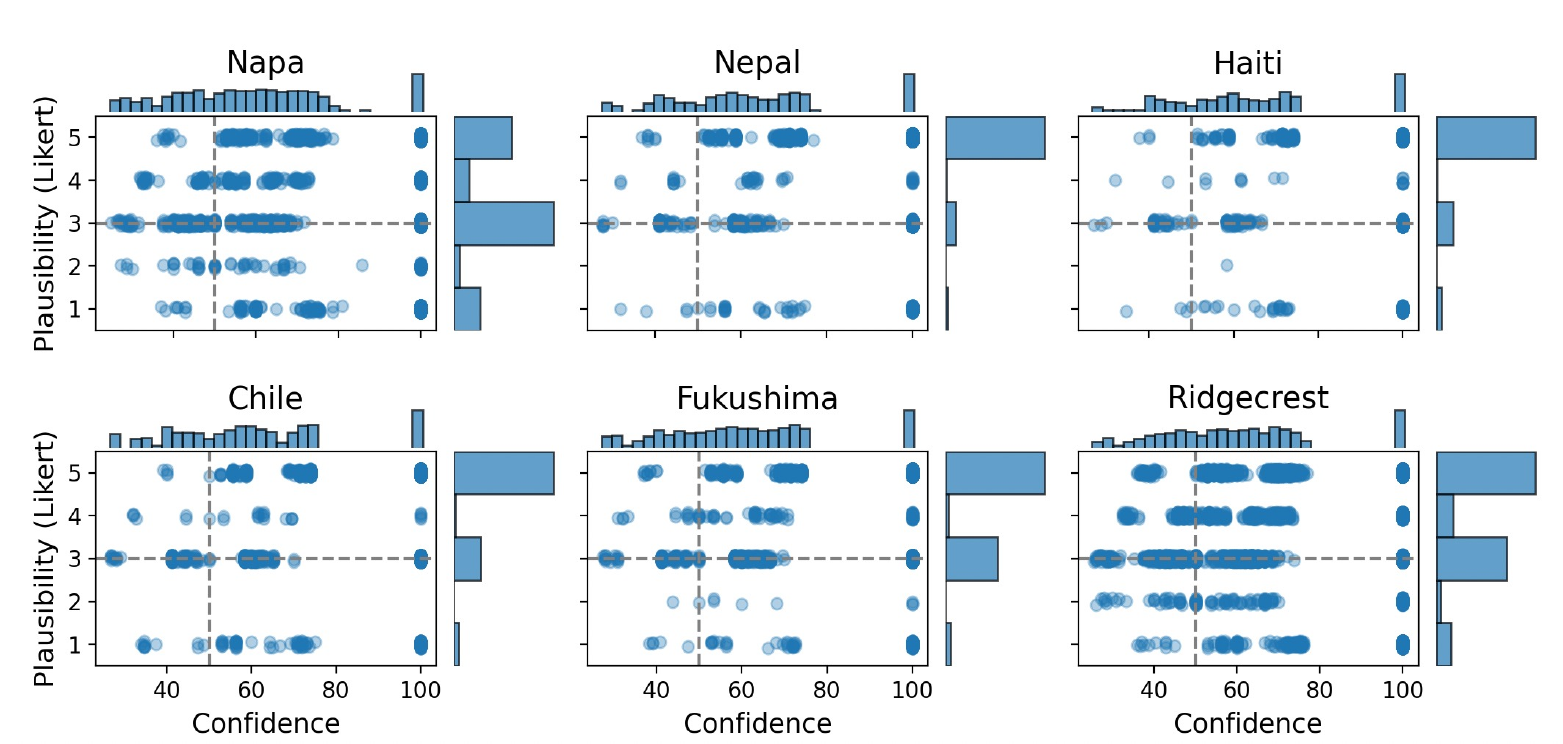} \caption{Joint distributions of plausibility and confidence estimates for six earthquake events. Each panel shows a scatter plot of tweets with marginal histograms above (confidence, log scale) and to the right (plausibility). Dashed lines indicate quadrant boundaries at confidence=50 and plausibility=3.} \label{fig:uncertainty_dist} 
\end{figure}

\paragraph{RQ1: Differentiation of situational signals}

We first examine whether the uncertainty-aware pipeline differentiates between tweet situational signals in terms of their epistemic uncertainty. Under a baseline approach, all SA-classified tweets are treated as equally plausible, and variation in uncertainty is not represented. In contrast, the uncertainty-aware pipeline assigns graded plausibility and confidence estimates to each tweet, explicitly modeling differences in epistemic status.

Figure \ref{fig:uncertainty_dist} illustrates the joint distribution of plausibility and confidence estimates for each earthquake in our case study. All events exhibit clear heterogeneity. Plausibility scores span the full five-point Likert scale, while confidence values range from approximately 20\% to 100\%. Although a substantial concentration of assignments fall at maximal confidence (100\%), dispersion persists across lower confidence levels, indicating that the model does not assign uniform certainty to all claims.

The concentration of maximal confidence values is consistent with prior research showing that LLMs tend to exhibit overconfidence when explicitly verbalizing their certainty \cite{xiong2023can}. This tendency therefore reflects a known characteristic of confidence elicitation rather than a failure of the framework. Importantly, even with this bias toward high confidence, the model does not collapse situational signals into a single epistemic category. Variation remains across both plausibility and confidence dimensions, indicating that the uncertainty-aware pipeline introduces structured differentiation relative to a uniform baseline.

While the distributional plots in Figure \ref{fig:uncertainty_dist} provide a visual indication of epistemic variation, we next quantify this heterogeneity using Shannon entropy (Equation \ref{eq:entropy}). Entropy provides a scalar measure of dispersion, enabling systematic comparison across events and against the baseline where all tweets are assigned identical uncertainty values.

Table \ref{tab:epistemic_rq1} summarizes entropy statistics across earthquake events. Across all six case studies, both plausibility and confidence entropy are strictly greater than zero, in contrast to the baseline’s entropy of 0. This indicates that the uncertainty-aware framework introduces measurable dispersion in epistemic assessments rather than collapsing all tweets into a single category. The magnitude of dispersion varies by event. For example, the Napa case study exhibits particularly high plausibility entropy (1.82 bits, out of a theoretical maximum of 2.32 for a five-point scale), indicating substantial spread across the Likert scale. Its confidence entropy (1.21 bits) similarly reflects variability in the strength of plausibility judgments.

Together, the distributional and entropy analyses provide complementary evidence for RQ1: the proposed framework introduces structured epistemic differentiation into tweet-level inputs that would otherwise remain undifferentiated under baseline aggregation.

\begin{table}
\centering
\small
\setlength{\tabcolsep}{3pt}
\caption{Entropy values for uncertainty-aware inputs across earthquake use cases.}
\label{tab:epistemic_rq1}
\begin{tabular}{lccc}
\toprule
\textbf{Event} & \textbf{Num. Tweets} & \textbf{Likert Entropy (bits)} & \textbf{Confidence Entropy (bits)} \\
\midrule
\textit{Baseline} & -- & 0.00 & 0.00 \\
Ridgecrest 2019 & 8,725 & 0.91 & 0.71 \\
Nepal 2015 & 6,201 & 0.95 & 0.67 \\
Haiti 2021 & 2,604 & 0.89 & 0.57 \\
Napa 2014 & 3,639 & 1.82 & 1.21 \\
Chile 2014 & 3,540 & 1.00 & 0.89 \\
Fukushima 2021 & 3,972 & 1.25 & 0.87 \\
\bottomrule
\end{tabular}
\end{table}

\paragraph{RQ2: Differentiation in generated reports}

While RQ1 evaluates epistemic differentiation at the level of individual situational signals, RQ2 examines whether this structure propagates into downstream report generation. To assess this, we analyze two complementary report-level metrics: (i) cross-report semantic similarity, which measures how distinct uncertainty-conditioned reports are from one another and from the baseline summary (Equation \ref{eq:report_similarity}), and (ii) internal semantic consistency, which measures coherence within each report (Equation \ref{eq:report_internal_consistency}).

Cross-report similarity is the primary test of downstream impact. If reports generated from different epistemic regimes remain highly similar to the baseline and to each other, then tweet-level uncertainty differentiation may not meaningfully alter the composition of information. In contrast, lower similarity might indicate conditioning on plausibility and confidence leads to substantively different summaries.

\begin{table}[t!]
\centering
\small
\caption{Evaluation metrics for all reports across events. \textit{IC} refers to internal consistency while \textit{Sim.} refers to similarity across reports.}
\label{tab:rq2_results}
\setlength{\tabcolsep}{3pt} 
\begin{tabular}{lcc|cc|cc|cc|cc|cc}
\toprule
 & \multicolumn{2}{c}{\textbf{Napa}} & \multicolumn{2}{c}{\textbf{Chile}} & \multicolumn{2}{c}{\textbf{Fukushima}} & \multicolumn{2}{c}{\textbf{Haiti}} & \multicolumn{2}{c}{\textbf{Ridgecrest}} & \multicolumn{2}{c}{\textbf{Nepal}} \\
\textbf{Report} & IC & Sim. & IC & Sim. & IC & Sim. & IC & Sim. & IC & Sim. & IC & Sim. \\
\midrule
Q1 & 0.31 & 0.83 & 0.32 & 0.86 & 0.31 & 0.85 & 0.30 & 0.83 & 0.22 & 0.87 & 0.29 & 0.82 \\
Q2 & 0.29 & 0.72 & 0.34 & 0.78 & 0.28 & 0.78 & 0.29 & 0.83 & 0.18 & 0.83 & 0.29 & 0.74 \\
Q3 & 0.26 & 0.75 & 0.33 & 0.71 & 0.30 & 0.71 & 0.26 & 0.73 & 0.24 & 0.82 & 0.25 & 0.71 \\
Q4 & 0.41 & 0.72 & 0.58 & 0.91 & 0.28 & 0.65 & 0.30 & 0.86 & 0.34 & 0.77 & 0.29 & 0.67 \\
Baseline & 0.31 & 0.85 & 0.29 & 0.88 & 0.31 & 0.87 & 0.32 & 0.85 & 0.26 & 0.88 & 0.29 & 0.86 \\
\bottomrule
\end{tabular}
\end{table}

\begin{figure}
    \centering
    \includegraphics[width=12cm]{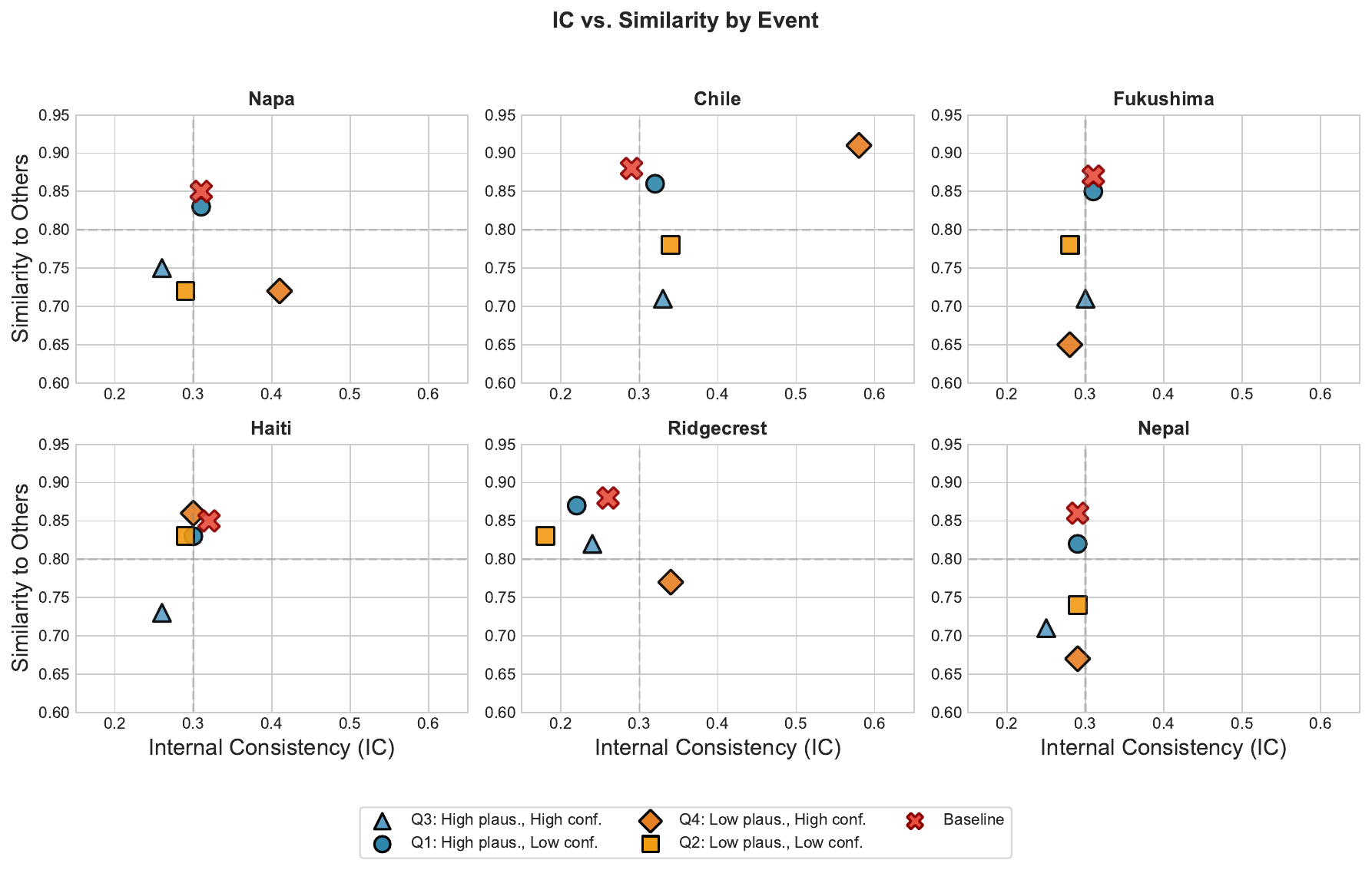}
    \caption{Internal consistency (IC) versus cross-report similarity for each event, with points colored by report type (Q1--Q4 based on plausibility and confidence estimates (plus baseline).}
    \label{fig:res_sim}
\end{figure}

Table \ref{tab:rq2_results} reports these metrics across all earthquake events, while Figure \ref{fig:res_sim} visualizes their relationship. Pairwise similarity scores show variation across quadrant-specific reports. For example, in Fukushima, Q4 exhibits the lowest similarity (0.65), indicating that this report differs noticeably from the others. However, this pattern is not uniform across events. In some cases, quadrant-specific reports remain highly similar to the baseline, potentially reflecting shared dominant themes within the underlying tweet corpus.

Internal consistency provides a complementary diagnostic. Across several events, quadrant-specific reports exhibit higher internal coherence than the baseline. For instance, in the Chile event, Q4 achieves an internal consistency of 0.58 compared to 0.29 for the baseline, while in Napa Q4 reaches 0.41 versus 0.31. This suggests that grouping tweets by plausibility and confidence might produce more internally focused summaries, even when overall thematic overlap remains.

Taken together, these findings indicate that uncertainty assessment does not merely partition tweets arbitrarily. In multiple cases, epistemic structuring produces semantically differentiated summaries, demonstrating that tweet-level uncertainty estimates can propagate into observable changes in report composition. At the same time, persistent similarity in some settings highlights that dominant crisis themes may constrain the extent of downstream divergence.
\section{Discussion}\label{sec:discussion}

\subsection{Interpreting the uncertainty-aware framework}

The results demonstrate that the proposed framework introduces epistemic differentiation at the tweet level (RQ1) and, in multiple cases, propagates this structure into distinct downstream reports (RQ2). At the same time, the distribution of tweets across plausibility-confidence quadrants is not uniform. In several events, certain quadrants are sparsely populated, and not all grid cells contain sufficient tweets to generate all quadrant-specific summaries.

This sparsity should not be interpreted as a limitation of the pipeline. The framework does not enforce balanced epistemic partitions; instead, tweets are assigned to quadrants based on plausibility and confidence thresholds, meaning some partitions may contain substantially fewer observations. If no tweets fall into a given quadrant, this reflects the informational landscape of the event rather than a system failure. In this sense, the pipeline functions diagnostically; it reveals the epistemic distribution present in the data rather than imposing artificial symmetry.

Across events, the least frequently populated quadrant is Q2 (low plausibility, low confidence). This profile corresponds to claims judged implausible and weakly supported by contextual evidence. Its relative rarity suggests that crisis-related discourse rarely occupies this combined epistemic position. Implausible claims may often be expressed with rhetorical certainty, placing them in Q4 (low plausibility, high confidence), while ambiguous or emerging claims may remain plausible but associated with low confidence (Q1). Alternatively, the model may preferentially express uncertainty through reduced confidence rather than simultaneously lowering both plausibility and confidence. Further investigation would be required to disentangle these explanations.

More broadly, the RQ2 findings clarify the operational value of uncertainty-aware structuring. In several events, conditioning on epistemic profiles led to reports that differed measurably from baseline summaries. This indicates that modeling uncertainty can affect not only how individual claims are characterized, but also how the overall situation is described. However, persistent similarity in some cases highlights that dominant crisis themes may constrain downstream divergence. The framework therefore does not guarantee radically different narratives. Instead, it provides structured pathways through which epistemic variation can influence summary generation when present.

\subsection{Operational use in crisis communication}

The proposed framework is designed to support, rather than replace, human decision-makers involved in time-critical disaster response and crisis communication. In operational settings such as emergency operations centers (EOCs), humanitarian response teams, or public information offices, analysts are often required to rapidly synthesize large volumes of social media reports while accounting for uncertainty and misinformation. The quadrant-based structuring of situational signals provides a mechanism for epistemic triage that can assist this process. To illustrate its intended use, consider the unfolding of a major earthquake in a densely populated region.

\paragraph{Early phase: minutes to hours after impact.}

Within minutes of the earthquake, social media platforms (such as X) may begin to fill with reports of shaking intensity, infrastructure damage, and potential requests for assistance. At this stage, information may be highly fragmented and uncertain. Some claims may align with emerging seismic data or official updates, while others may reflect confusion, rumor, or localized misinterpretation.

In this context, the uncertainty assessment module structures incoming situational signals into epistemically differentiated subsets. High-plausibility, high-confidence (Q3) reports can support rapid internal briefings, helping analysts form an initial operational picture. High-plausibility, low-confidence (Q1) reports may indicate emerging developments (such as reports of bridge damage or power outages) that warrant monitoring or targeted verification. Low-plausibility, high-confidence (Q4) reports may signal confidently expressed claims that diverge from proxy indicators. While such cases can reflect rumor propagation or misinformation, they may also correspond to localized outlier events not captured in coarse-grained impact summaries. For example, during the March 2025 7.7 magnitude earthquake in Myanmar, Thai authorities confirmed that a single high-rise building in Bangkok collapsed due to tremors from the distant quake, despite limited structural damage elsewhere in the city \cite{reuters_thailand_2025}. In a proxy-informed framework relying on aggregate regional impact indicators, such a localized event might initially appear implausible while being reported with high confidence. This illustrates that divergence from coarse-grained proxy estimates does not necessarily imply falsehood, but may instead reflect spatially concentrated outlier events. Low-plausibility, low-confidence reports, by contrast, are more likely to reflect ambiguous or weakly supported content that can be deprioritized during early response. Rather than producing a single aggregated summary of all tweets, the system therefore provides structured informational views that support decision-making under uncertainty.

\paragraph{Stabilization phase: hours to days after impact.}

As official, verified reports from trusted data sources accumulate, the informational environment arguably becomes more stable. In this phase, quadrant reports may serve different functions. High-plausibility, high-confidence (Q3) reports may support public communication updates or coordination with partner agencies. Previously low-confidence but plausible (Q1) claims can be revisited as additional corroborating evidence emerges. Conversely, persistently low-plausibility, high-confidence (Q4) narratives may warrant monitoring for misinformation risks or public reassurance campaigns.

We emphasize here that this framework is not a fact-checking tool, nor is it intended to replace human expertise. Instead, it makes explicit the uncertainty inherent in social media situational signals by conditioning plausibility assessments on external proxy data. Human analysts remain responsible for interpretation, validation, and external communication, while the system assists in structuring large volumes of situational data into manageable and epistemically differentiated reports.
\section{Limitations and future work}\label{sec:limitations}

There are several limitations to this work. First, although the uncertainty assessment framework is designed to be adaptable to different crisis types, our empirical evaluation focuses only on earthquakes. In this study, plausibility is assessed using proxy data from the USGS PAGER system. Comparable proxy datasets may not be readily available for all hazard types. Applying the framework to events such as hurricanes, floods, or wildfires would therefore require identifying alternative external data sources that provide structured and geographically aligned impact information, consistent with the requirements outlined in Section \ref{sec:methodology_proxy}.

Second, quadrant-based report generation depends on the choice of plausibility and confidence thresholds. In this study, we used neutral split points (Likert $=3$ for plausibility and confidence $=50\%$) to divide tweets into uncertainty profiles. These values were not tuned or optimized. Different threshold choices would change how tweets are distributed across quadrants and could influence the resulting summaries. A systematic sensitivity analysis of alternative threshold configurations is left for future work.

Third, our evaluation relies on embedding-based similarity measures to assess internal consistency and differences between reports. These metrics provide a scalable way to compare report structure and content, but they do not directly measure factual accuracy or real-world usefulness. Human evaluation with domain experts would provide stronger evidence of operational value. Assessing factual correctness, however, was beyond the scope of this study.

Fourth, the framework's modular design --- separating classification, uncertainty assessment, and report generation --- reflects a deliberate attempt to manage prompt complexity. One alternative could be an agentic workflow, where a single autonomous agent, for example, iteratively reasons about classification, plausibility, and summarization within a unified prompt structure. Such agentic designs can enable dynamic verification between sub-tasks and may reduce prompt overhead by internalizing hierarchical reasoning. However, in high-stakes crisis contexts, separating tasks reduces prompt complexity and enhances interpretability for analysts. Nevertheless, exploring agentic alternatives, and their effects on output quality, remains a promising avenue for future research. 

\section{Conclusion}\label{sec:conclusion}

This paper has introduced an LLM-based uncertainty assessment framework for social media situational signals designed to support crisis reporting. While recent automated crisis reporting systems have shown LLMs can synthesize situational posts into coherent summaries, they arguably rely on the implicit assumption that all posts are equally plausible. We address this limitation by designing a framework that evaluates claims against external proxy data to produce plausibility judgments, while also eliciting the model's self-expressed confidence in those judgments. This transforms raw situational signals into tweets with explicit plausibility and confidence estimates, enabling reports that reflect not only what is being claimed, but also how certain those claims appear given available proxy information.

Across six earthquake case studies, we demonstrate how incorporating uncertainty assessment produces meaningful variation in tweet-level plausibility and confidence estimates, challenging the baseline assumption that all classified signals are equally plausible. This differentiation propagates into downstream report generation, allowing comparison between reports generated from tweets with different uncertainty profiles. Rather than ignoring uncertainty, the proposed framework models it explicitly to distinguish between more and less plausible situational signals. This helps decision makers prioritize information when interpreting large volumes of social media data in real-time, enabling more transparent automated reporting at scale.

\backmatter

\section*{Funding}
Timothy Douglas is supported by the EPSRC DTP Research Studentship provided by University College London. This work was conducted during the internships of Timothy Douglas and Roben Delos Reyes at the National Institute of Informatics (NII) and was supported by the NII International Internship Program.

\section*{Acknowledgments}
The experiments conducted in this work were supported by the resources provided by University College London's Research Computing Services. We would also like to thank Lingyao Li for sharing the dataset used in this work.

\section*{Code availability}

The source code used in this study is available at: \url{https://github.com/TimDouglas28/uncertainty_assessment}

\bibliographystyle{unsrt}
\bibliography{sn-bibliography}
\clearpage
\appendix

\noindent{\LARGE Supplementary Material \par}

\subsection*{Classification Prompt}

\begin{promptbox}[Box 1: Classification Prompt]{gray}
\textbf{System role.}  
You are an information extraction system assisting earthquake impact researchers.
\\[6pt]

\textbf{Task.}  
Classify each tweet into one or more situational awareness categories and extract the key text spans (``rationales'') that justify each classification.
\\[6pt]

\textbf{Situational awareness schema.}
\\
\V{SITUATIONAL AWARENESS SCHEMA}
\\[6pt]

\textbf{Input.}  
Now read the following social media posts:
\\
\V{SOCIAL MEDIA POSTS}
\\[6pt]

\textbf{Output requirements.}
\begin{itemize}
    \item Return \textbf{only} a valid JSON array.
    \item Include \textbf{all} fields for each tweet:
    \texttt{index}, \texttt{tweet\_id}, \texttt{tweet\_text},
    \texttt{situational\_categories}, \texttt{rationales}.
    \item Assign all relevant categories (zero, one, or multiple).
    \item Use \textbf{only} the exact category names listed above
    (lowercase with underscores).
    \item For \texttt{rationales}, include specific text snippets from the tweet that support \textbf{each} assigned category.
\end{itemize}

\textbf{Required JSON format (per tweet).}
\begin{verbatim}
{
  "index": <original_index_number>,
  "tweet_id": <unique_tweet_identifier>,
  "tweet_text": <full_text_of_tweet>,
  "situational_categories": ["<category1>", "<category2>", ...],
  "rationales": ["<key_phrase_1>", "<key_phrase_2>", ...]
}
\end{verbatim}

\textbf{Critical formatting rules.}
\begin{enumerate}
    \item Do \textbf{not} use markdown code blocks (e.g., \texttt{```json}).
    \item Do \textbf{not} add explanations, headers, or separators (e.g., ``---'').
    \item The JSON must start with \texttt{[} and end with \texttt{]}.
    \item Do \textbf{not} wrap the JSON in any other text or formatting.
    \item Do \textbf{not} include example output or placeholder text.
\end{enumerate}
\end{promptbox}

\newpage
\thispagestyle{empty}
\subsection*{Uncertainty Assessment Prompt}
\begin{promptbox}[Box 2: Uncertainty Assessment Prompt]{gray}
\textbf{System role.}  
You are an expert crisis analysis assistant.
\\[6pt]

\textbf{Task.}  
Evaluate whether social media situational awareness (SA) signals
are plausibly representative of real-world conditions during a crisis event, given automatically generated impact estimates and contextual uncertainty.
\\[6pt]

\textbf{Context.}
\\
Event: \V{EVENT NAME}
\\
Region: \V{MMI}
\\
Time: \V{TIME}
\\

\textbf{External Proxy Data.}
\\
Proxy Summary: \V{SUMMARY}
\\[6pt]

\textbf{Input.}  
Now read the following social media posts:
\\
\V{SOCIAL MEDIA POSTS}
\\[6pt]

\textbf{Uncertainty assessment task.}  
For \emph{each tweet}, determine the following:

\begin{enumerate}
    \item \textbf{Plausibility (Likert 1–5).}  
    To what extent could this tweet plausibly reflect a real-world situational development during this event, given:
    \begin{itemize}
        \item The external impact proxy.
        \item Known limitations of the proxy.
        \item The specificity, tone, timing, and credibility of the tweet content.
    \end{itemize}
    \emph{Note: A tweet may be plausible even if it contradicts the proxy.}
    \begin{itemize}
        \item 5 = highly plausible
        \item 3 = uncertain or ambiguous
        \item 1 = implausible
    \end{itemize}

    \item \textbf{Confidence (0–100).}  
    How confident are you in \emph{your plausibility judgment}?  
    This reflects uncertainty in the judgment itself, not confidence that the proxy is correct.

    \item \textbf{Discrepancy assessment.}  
    If the tweet appears weakly plausible or implausible, indicate the most likely reason:
    \begin{itemize}
        \item \texttt{misclassification} — the assigned SA category may be incorrect.
        \item \texttt{noise} — the tweet is vague, exaggerated, irrelevant, or unreliable.
        \item \texttt{none} — no clear discrepancy.
    \end{itemize}

    \item \textbf{Rationale.}  
    Briefly explain your reasoning, explicitly referencing:
    \begin{itemize}
        \item The tweet content and language.
        \item The assigned SA category or categories.
        \item The timing of the tweet.
        \item How the claim relates to or diverges from the proxy context.
        \item Why the claim may still be plausible (or not).
    \end{itemize}
\end{enumerate}
\vspace{6pt}
\textbf{Output requirements.} \textit{Similar to Box 1}

\textbf{Required JSON format (per tweet).}
\begin{verbatim}
{
  "index": <original_index_number>,
  "tweet_id": <unique_tweet_identifier>,
  "situational_categories": ["<category1>", "<category2>", ...],
  "likert_plausibility": <integer 1–5>,
  "confidence": <integer 0–100>,
  "discrepancy_assessment": "<misclassification | noise | none>",
  "reason_alignment": "<brief explanation>"
}
\end{verbatim}

\textbf{Critical formatting rules.} \textit{Identical to Box 1}
\end{promptbox}

\clearpage
\subsection*{Real-world proxy data extracted from PAGER}

The following tables present the complete PAGER proxy data extracted for each earthquake case study, including perceived shaking, estimated potential damage for resistant and vulnerable structures, and population exposure at each MMI level. Population exposure values marked with ``-'' indicate no data reported. Data sourced from USGS PAGER system summaries.

\begin{table}[htbp]
\centering
\caption{PAGER proxy data for the 2014 Iquique earthquake.}
\label{tab:pager_iquique}
\begin{tabular}{c|c|c|c|c}
    \toprule
    \textbf{MMI} & \textbf{Perceived} & \textbf{Damage} & \textbf{Damage} & \textbf{Population} \\
    & \textbf{Shaking} & \textbf{(Resistant)} & \textbf{(Vulnerable)} & \\
    \midrule
    I & Not felt & None & None & - \\
    II & Weak & None & None & - \\
    III & Weak & None & None & - \\
    IV & Light & None & None & 360k \\
    V & Moderate & Very Light & Light & 240k \\
    VI & Strong & Light & Moderate & 282k \\
    VII & Very Strong & Moderate & Moderate/Heavy & 501k \\
    VIII & Severe & Moderate/Heavy & Heavy & 0 \\
    IX & Violent & Heavy & Very Heavy & 0 \\
    X & Extreme & Very Heavy & Very Heavy & 0 \\
    \bottomrule
\end{tabular}
\end{table}

\begin{table}[htbp]
\centering
\caption{PAGER proxy data for the 2014 Napa earthquake.}
\label{tab:pager_napa}
\begin{tabular}{c|c|c|c|c}
    \toprule
    \textbf{MMI} & \textbf{Perceived} & \textbf{Damage} & \textbf{Damage} & \textbf{Population} \\
    & \textbf{Shaking} & \textbf{(Resistant)} & \textbf{(Vulnerable)} & \\
    \midrule
    I & Not felt & None & None & - \\
    II & Weak & None & None & 4,881k \\
    III & Weak & None & None & 4,881k \\
    IV & Light & None & None & 3,281k \\
    V & Moderate & V. Light & Light & 370k \\
    VI & Strong & Light & Moderate & 145k \\
    VII & Very Strong & Moderate & Moderate/Heavy & 52k \\
    VIII & Severe & Moderate/Heavy & Heavy & 82k \\
    IX & Violent & Heavy & Very Heavy & 0 \\
    X & Extreme & Very Heavy & Very Heavy & 0 \\
    \bottomrule
\end{tabular}
\end{table}

\begin{table}[htbp]
\centering
\caption{PAGER proxy data for the 2015 Nepal earthquake.}
\label{tab:pager_nepal}
\begin{tabular}{c|c|c|c|c}
    \toprule
    \textbf{MMI} & \textbf{Perceived} & \textbf{Damage} & \textbf{Damage} & \textbf{Population} \\
    & \textbf{Shaking} & \textbf{(Resistant)} & \textbf{(Vulnerable)} & \\
    \midrule
    I & Not felt & None & None & - \\
    II & Weak & None & None & - \\
    III & Weak & None & None & - \\
    IV & Light & None & None & 10,721k \\
    V & Moderate & Very Light & Light & 84,253k \\
    VI & Strong & Light & Moderate & 40,899k \\
    VII & Very Strong & Moderate & Moderate/Heavy & 3,556k \\
    VIII & Severe & Moderate/Heavy & Heavy & 2,885k \\
    IX & Violent & Heavy & Very Heavy & 12k \\
    X & Extreme & Very Heavy & Very Heavy & 0 \\
    \bottomrule
\end{tabular}
\end{table}

\begin{table}[htbp]
\centering
\caption{PAGER proxy data for the 2019 Ridgecrest earthquake.}
\label{tab:pager_ridgecrest}
\begin{tabular}{c|c|c|c|c}
    \toprule
    \textbf{MMI} & \textbf{Perceived} & \textbf{Damage} & \textbf{Damage} & \textbf{Population} \\
    & \textbf{Shaking} & \textbf{(Resistant)} & \textbf{(Vulnerable)} & \\
    \midrule
    I & Not felt & None & None & - \\
    II & Weak & None & None & 28,545k \\
    III & Weak & None & None & 28,545k \\
    IV & Light & None & None & 21,546k \\
    V & Moderate & Very Light & Light & 602k \\
    VI & Strong & Light & Moderate & 2k \\
    VII & Very Strong & Moderate & Moderate/Heavy & 45k \\
    VIII & Severe & Moderate/Heavy & Heavy & 0 \\
    IX & Violent & Heavy & Very Heavy & 0 \\
    X & Extreme & Very Heavy & Very Heavy & 0 \\
    \bottomrule
\end{tabular}
\end{table}

\begin{table}[htbp]
\centering
\caption{PAGER proxy data for the 2021 Fukushima earthquake.}
\label{tab:pager_fukushima}
\begin{tabular}{c|c|c|c|c}
    \toprule
    \textbf{MMI} & \textbf{Perceived} & \textbf{Damage} & \textbf{Damage} & \textbf{Population} \\
    & \textbf{Shaking} & \textbf{(Resistant)} & \textbf{(Vulnerable)} & \\
    \midrule
    I & Not felt & None & None & - \\
    II & Weak & None & None & 28,521k \\
    III & Weak & None & None & 28,521k \\
    IV & Light & None & None & 36,653k \\
    V & Moderate & V. Light & Light & 15,388k \\
    VI & Strong & Light & Moderate & 1,862k \\
    VII & Very Strong & Moderate & Moderate/Heavy & 2,308k \\
    VIII & Severe & Moderate/Heavy & Heavy & 105k \\
    IX & Violent & Heavy & Very Heavy & 0 \\
    X & Extreme & Very Heavy & Very Heavy & 0 \\
    \bottomrule
\end{tabular}
\end{table}

\begin{table}[t!]
\centering
\caption{PAGER proxy data for the 2021 Haiti earthquake.}
\label{tab:pager_haiti}
\begin{tabular}{c|c|c|c|c}
    \toprule
    \textbf{MMI} & \textbf{Perceived} & \textbf{Damage} & \textbf{Damage} & \textbf{Population} \\
    & \textbf{Shaking} & \textbf{(Resistant)} & \textbf{(Vulnerable)} & \\
    \midrule
    I & Not felt & None & None & - \\
    II & Weak & None & None & 3,001k \\
    III & Weak & None & None & 3,001k \\
    IV & Light & None & None & 12,482k \\
    V & Moderate & Very Light & Light & 5,032k \\
    VI & Strong & Light & Moderate & 804k \\
    VII & Very Strong & Moderate & Moderate/Heavy & 937k \\
    VIII & Severe & Moderate/Heavy & Heavy & 340k \\
    IX & Violent & Heavy & Very Heavy & 0 \\
    X & Extreme & Very Heavy & Very Heavy & 0 \\
    \bottomrule
\end{tabular}
\end{table}\label{sec:app}

\end{document}